\documentclass[sigconf, authorversion=true]{acmart}

\AtBeginDocument{%
  \providecommand\BibTeX{{%
    \normalfont B\kern-0.5em{\scshape i\kern-0.25em b}\kern-0.8em\TeX}}}

\copyrightyear{2020} 
\acmYear{2020} 
\setcopyright{acmlicensed}
\acmConference[MSR '20]{17th International Conference on Mining Software Repositories}{October 5--6, 2020}{Seoul, Republic of Korea}
\acmBooktitle{17th International Conference on Mining Software Repositories (MSR '20), October 5--6, 2020, Seoul, Republic of Korea}
\acmPrice{15.00}
\acmDOI{10.1145/3379597.3387487}
\acmISBN{978-1-4503-7517-7/20/05}


\usepackage[utf8]{inputenc}
\usepackage{url}
\usepackage{flushend}
\usepackage{graphicx}

\usepackage{xspace}
\newcommand{\etal}{et al.\xspace}
\newcommand{\ie}{i.e.,\xspace}
\newcommand{\eg}{e.g.,\xspace}

\newcommand{\fig}[1]{Fig.~\ref{#1}}
\newcommand{\tab}[1]{Table~\ref{#1}}

\newcommand{\sect}[1]{Section~\ref{#1}}

\begin{document}

\title{20-MAD - 20 Years of Issues and Commits of\\Mozilla and Apache Development}

\author{Maëlick Claes}
\affiliation{M3S, ITEE, University of Oulu, Finland}
\email{maelick.claes@oulu.fi}
\orcid{0000-0003-2259-3946}

\author{Mika V. Mäntylä}
\affiliation{M3S, ITEE, University of Oulu, Finland}
\email{mika.mantyla@oulu.fi}

\begin{abstract}
Data of long-lived and high profile projects is valuable for research on successful software engineering in the wild. Having a dataset with different linked software repositories of such projects, enables deeper diving investigations. This paper presents 20-MAD, a dataset linking the commit and issue data of Mozilla and Apache projects. It includes over 20 years of information about 765 projects, 3.4M commits, 2.3M issues, and 17.3M issue comments, and its compressed size is over 6 GB.
The data contains all the typical information about source code commits (\eg lines added and removed, message and commit time) and issues (status, severity, votes, and summary). The issue comments have been pre-processed for natural language processing and sentiment analysis. This includes emoticons and valence and arousal scores. Linking code repository and issue tracker information, allows studying individuals in two types of repositories and provide more accurate time zone information for issue trackers as well.
To our knowledge, this the largest linked dataset in size and in project lifetime that is not based on GitHub.

\end{abstract}

\maketitle


\section{Introduction}

Data of long-lived and high profile projects is valuable for research on successful software engineering in the wild. Having a dataset with different linked software repositories of such projects, enables deeper diving investigations.
Yet, linking data from different software repositories, such as code
repositories and issue trackers, for mining purpose can be
challenging. During the past decade, GitHub has been a popular way for
accessing linked data about code commits and issue/bug information.
However, it can be common for large and long-lived open
source projects, such as Mozilla and Apache, to rely on their own
issue trackers rather than GitHub's, even when the projects provide
GitHub mirrors.

In this paper, we presents 20-MAD, a dataset linking the commit and issue data of Mozilla and Apache projects. It includes over 20 years of information about 765 projects, 3.4M commits, 2.3M issues, and 17.3M issue comments. For comparison Ortu \etal~\cite{OrtuJira} dataset has 700K issues and 2M issue comments but no commit information. Commits and issues are linked in two ways: user ids belonging to the same developer\footnote{For data protection, user personal information is anonymized in the dataset.} in code repositories and issue trackers, and through issue id that is available in over 80\% of the commit messages.

In addition to providing linked meta-data about commits and issues, the dataset was processed in various ways:
\begin{itemize}
    \item Semi-automatic identity merging of developer profiles.
    \item Timezone information extracted from commit repository and matched with issue tracker timestamps.
    \item Natural language processing of issue tracker comments including, filtering out of source code with NLoN~\cite{mantyla2018nlon}, sentence tokenization with the R package \emph{tokenizers}, emoticon extraction with our own R package\footnote{\url{https://github.com/M3SOulu/EmoticonFindeR}} and sentiment analysis using SentiStrength~\cite{sentistrength} and Senti4SD~\cite{Calefato2018}.
\end{itemize}

This dataset originates from our MSR 2017 study on abnormal work hours in Mozilla Firefox~\cite{claes2017abnormal} for which we extracted Mozilla commit and issue tracker information. We updated it and extended it with Apache code repositories in order to compare work hours in Mozilla projects for an ICSE 2018 publication~\cite{Claes2018ICSE}. We also used it for our MSR 2018 study on paid developers~\cite{claes2018towards}. Meanwhile, we extended the dataset further with issue tracker information in order to perform natural language processing of comments. While we initially considered reusing the Jira dataset from Ortu \etal~\cite{OrtuJira}, because of missing accurate timezone information in older Apache commits, we re-extracted Apache's Jira data for our study of emoticons usage by Mozilla and Apache developers~\cite{Claes2018emoticons}.

The dataset presented here is an updated version of the dataset we used for our previous studies. It includes open source software development information spanning more than 20 years, going back from January 2020 to 1998 (commits) and 1994 (issues) for Mozilla and 1998 (commits) and 2003 (issues) for Apache. To the best of our knowledge, this is the largest dataset that links commit and issue tracker information without relying on GitHub. In addition, we are not aware of any other dataset providing time zone information in issue trackers. Finally, processing all the issue tracker comments with NLP tools such as NLoN and Senti4SD would take around 20 days of computations with an average laptop.

The dataset can be used for various repository mining tasks or studies that would require access to developers weekly and daily work patterns, large scale sentiment analysis, analysis of emoticon usage by software developers or NLP tasks of comments. It can also be used to create a software engineering language model that isn't based on StackOverflow, contrarily to existing ones~\cite{efstathiou2018word}. While the dataset does not provide any information about source code, it can easily be extended with any other data that keeps track of commit hash in Apache or Mozilla projects.

Our dataset and source code are available in an Open Science Framework repository~\cite{data20mad, package20mad}. The documentation and source code are also available on GitHub\footnote{\url{https://github.com/M3SOulu/MozillaApacheDataset/releases/tag/msr2020}}.
Moreover, we provide a Docker image for simplifying the replication of processing the raw data\footnote{\url{https://hub.docker.com/repository/docker/claesmaelick/mozilla-apache-dataset}}.

In the remaining of this paper, we first present the information available in the dataset in \sect{sec:dataset}. Then, in \sect{sec:collection}, we provide details on how the data was extracted and processed. Finally, in \sect{sec:limitations}, we conclude by presenting the current limitations of the dataset that researchers should be aware of, and how we plan on improving it in the future.



\section{Description of the dataset}\label{sec:dataset}

\begin{figure}[t]
  \centering
  \includegraphics[width=0.9\columnwidth]{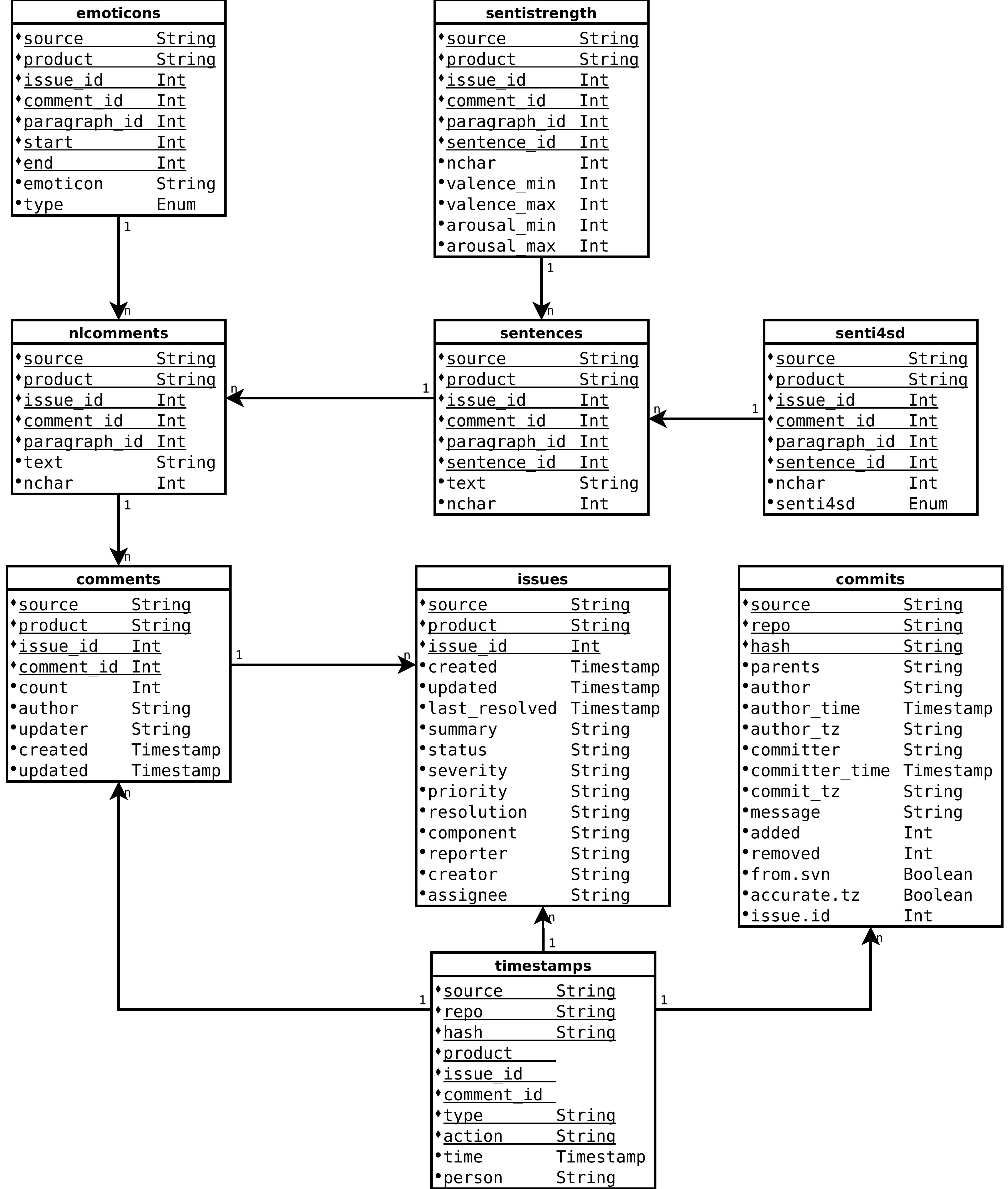}
  \caption{Simplified schema of the main Parquet files of the dataset. More information about the different fields can be found on the GitHub page of the dataset.}
  \label{fig:schema}
\end{figure}

The dataset consists of several tables. These are stored as Apache Parquet files\footnote{\url{https://parquet.apache.org/}} in order to keep column type information, such as timestamps.
There are two main sets of tables: logs and natural language
data. \fig{fig:schema} presents an overview of the dataset with the
different Parquet tables and their dependency relationships.

\begin{description}
\item[Logs] The logs contain meta-data information about commits,
  issues and issue comments:
  \begin{description}
  \item[Commits] Meta-data information about commits. A commit is
    u\-ni\-que\-ly identified by its source (\ie Apache or Mozilla), its
    repo(sitory) and its hash.
  \item[Issues] Meta-data information about issues. An issue is
    u\-ni\-que\-ly identified by its source, its
    product tag, and its id. Some fields are only available for
    issues coming from Jira or Bugzilla.
  \item[Comments] Meta-data information about issue com\-ments. A comment is uniquely identified by its sour\-ce (\ie Apache or Mozilla), its product tag, its issue id and its id.
  \item[Timestamps] Aggregated timestamp information from the three
    previous logs. A timestamp is uniquely identified by its source, its repository or product tag, its id\footnote{Hash for commits, issue id for issues and issue and commit
    ids for comments.} and its type and action.
  \end{description}
\item[NLP] We ran various natural language processing tools on the
  issue comments:
  \begin{description}
  \item[nlcomments] Part of comments that have been detected as
    natural language. Each comment is split by paragraphs and uniquely
    identified by its source, product tag, issue id, comment id and
    paragraph id.
  \item[emoticons] Emoticons and emojis found in natural language
    paragraphs. Each emoticon is uniquely identified by its source,
    product tag, issue id, comment id, paragraph id and its location in the paragraph. The type indicates whether it is
    a text emoticon or a Unicode emoji.
  \item[sentences] NL paragraphs are further split as
    sentences which are uniquely
    identified by source, product tag, issue id, comment id,
    paragraph id and sentence id.
  \item[sentistrength] Result of SentiStrength on each
    sentence using the default lexicon (min. and max. valence)
    and our lexicon~\cite{mantyla2017bootstrapping} for detecting arousal in software engineering
    context (min. and max. arousal).
  \item[senti4sd] Result of running Senti4SD on each sentence.
  \end{description}
\end{description}



\section{Data collection}\label{sec:collection}

\begin{figure*}[thp]
  \centering
  \includegraphics[width=0.8\textwidth]{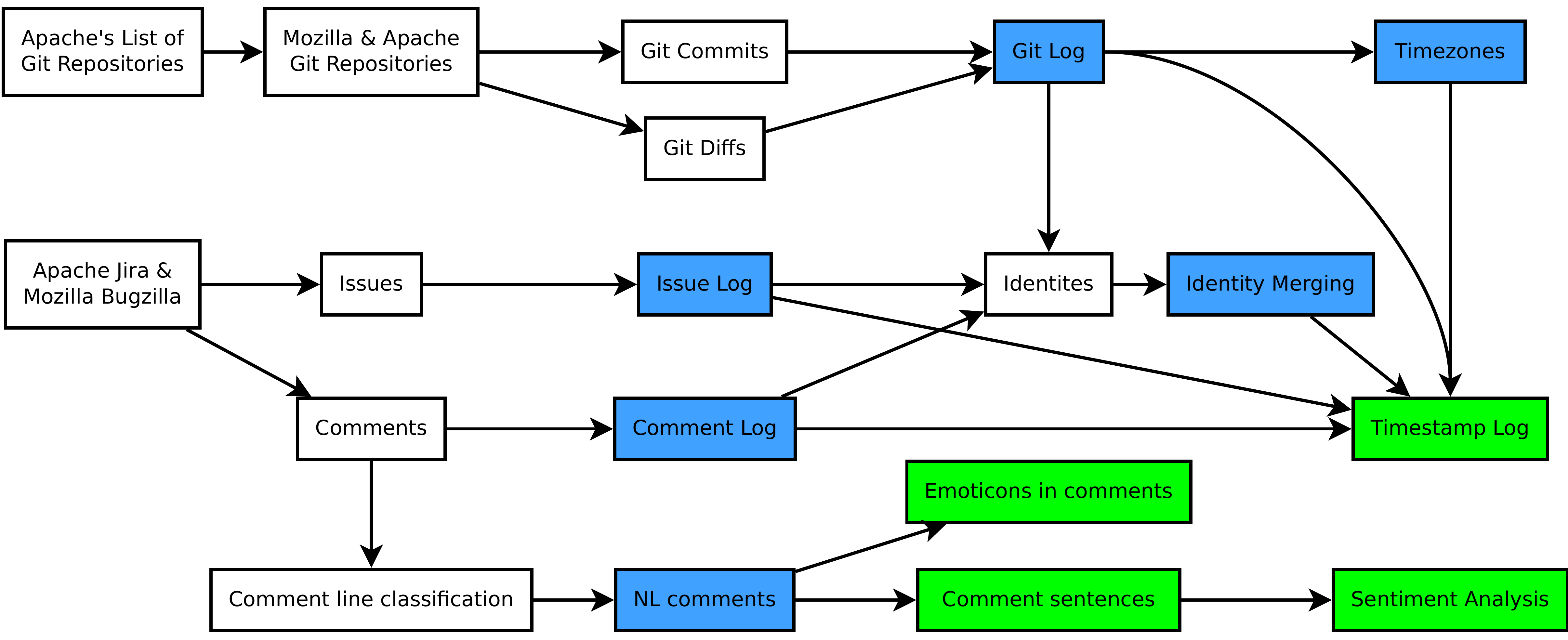}
  \caption{Dependency graph of the different steps for extracting and
    processing the data. Green rectangles represent the final
    processed data and blue rectangles pre-processed data that is also
    provided.}
  \label{fig:workflow}
\end{figure*}

In this section, we present how the data was collected and pre-processed. First, how the raw data from Git repositories and issue trackers was extracted using Perceval~\cite{DuenasPerceval} and the basic pre-processing applied to it. Second, we present the more elaborate processing steps including identity merging, time zone matching and natural language processing (NLP). \fig{fig:workflow} presents the workflow followed for data collection as a dependency graph.

\subsection{Data extraction}

The Git repositories were extracted between January 6th and 9th 2020. Mozilla's Bugzilla data was initially extracted in January 2017 and regularly updated until January 14th 2020. While the same was true for Apache's Jira, because of data corruption in 2019, it was completely re-extracted between January 14th and 20th 2020. Our estimations for processing all the data is around 20 days\footnote{On a laptop with 8GB of memory and an SSD drive.}. However, we relied on HPC resources\footnote{Provided by CSC (\url{https://wwww.csc.fi})}. The 16 days of computations estimated for running Senti4SD were reduced to less than 36 hours using 16 threads and up to 272GB of memory on the HPC environment.

\subsubsection{Git commits}

We parsed Apache's list of Git repositories\footnote{\url{https://issues.apache.org/jira}} using a custom-made Python script. This list provides the name and URL of the 1648\footnote{The number of Apache repository in the dataset is actually smaller due to empty Git repository listed on the website.} Git repositories currently maintained by the Apache community and groups them by projects. For Mozilla, we cloned the GitHub mirrors of the two main Mercurial repositories used by Mozilla: gecko-dev\footnote{\url{https://github.com/mozilla/gecko-dev}} and comm-central\footnote{\url{https://github.com/mozilla/releases-comm-central}}.

We cloned locally each repository and generated a log file containing all commits\footnote{With the following command: git log --raw --numstat --pretty=fuller --decorate=full --all --reverse --topo-order --parents -M -C -c -l100000 --remotes=origin}.
Each repository's generated commit log was parsed as a JSON file using Perceval~\cite{DuenasPerceval}. To make sure that each individual file stays small, we split the JSON file of Mozilla's gecko-dev into several smaller JSON files of 10,000 commits. Each JSON file was then converted to two Parquet tables: one containing the list of commits of the repository, and one containing the list of files changed in each commit. Time data (Git's commit and author dates) is also parsed as a timestamp and timezone.

\subsubsection{Issue trackers}

We used Perceval~\cite{DuenasPerceval} to extract issues from A\-pa\-che's Jira repository\footnote{\url{https://issues.apache.org/jira}} and
Mozilla's Bugzilla repository\footnote{\url{https://bugzilla.mozilla.org/home}} and stored the result in a MongoDB database. We exported the issues from the database into JSON files grouped by the product tag of each issue. Each product's JSON file was then further split into several files of maximum 10,000 issues.

Each JSON file was then converted to three or four Parquet tables: one containing the list of issues and associated meta-data, one containing the issue comments and associated meta-data, one containing the history of changes of the issues, and one containing the versions associated with the issues (only for Jira). The various time data associated with issues and comments (e.g. creation date, update date, resolved date) is also parsed as a timestamp and timezone.

\subsection{Data processing}

\subsubsection{Logs}

We produced several logs, lists of records of events, that aggregate the different Parquet tables generated before into a single one:
\begin{itemize}
    \item A commit log that merges commit log and diff by aggregating the total number of lines added and removed in each commit. In addition, issue tracker ids were extracted from commit messages in order to link commits and issues.
    \item An issue log that aggregates the issue meta-data from Jira and Bugzilla and fills the missing field with empty values.
    \item An issue comment log containing meta-data of all issue comments. The text of the comment is removed so metadata of all comments can fit as a single table in memory on a laptop with 8GB memory.
\end{itemize}

\subsubsection{Merging identities}



One of the goals of this dataset is to link commit and issue information. In order to make this possible, it is needed to link developer profiles used in code repositories and issue trackers. While issue trackers already have profiles that can be identified for each person, Git repositories only store a character string, usually formatted as \emph{Full Name {\textless}emailadress{\textgreater}}. Thus, it is common for a single developer to use multiple distinct identities.

We performed identity merging using a simple semi-automatic technique. We gathered the name and emails of all identities used in issue trackers (issue creators, updaters and assignees, and issue comment authors and updaters). For commits, we split the author and committer fields using regular expressions to obtain separated names and emails. For each identity gathered made of distinct lower-cased names and emails, we create a node in an undirected graph. Then we add edges between each node for which the name or the email is the same. Each connected component of the graph represents a distinct developer profile.

The result was manually verified and edges were manually added and removed to make sure that
\begin{itemize}
    \item distinct profiles were not merged together;
    \item the most active developers (in terms of number of commits) were properly linked to a profile in the issue tracker.
\end{itemize}
This manual verification was conducted for developers with more than 100 commits for the previous version of the data (extracted in January 2018 and used in our ICSE paper~\cite{Claes2018ICSE}), and re-checked for developers with more than 1000 commits for the current version. Overall, it increased the number of distinct (automatically merged) developer profiles by 198 for Mozilla and 258 for Apache.

This technique doesn't provide a perfect identity merging, yet we consider it to be good enough for studying the activity of regular developers in both issue and commit repositories. \tab{tab:idmerging} shows the numbers and percentages of developer profiles in Apache and Mozilla that have been linked to a profile in the associated issue tracker. We report a higher percentage for Mozilla than Apache as Mozilla developers have to open an issue in the Bugzilla repository in order to submit any code and get it reviewed. On the other hand, this is not required for Apache developers as individual projects can decide how to handle bug reports and code submissions.

\begin{table}[th]
  \centering
  \caption{Numbers and percentages of developer and issue tracker profiles.}
  \label{tab:idmerging}
  \begin{small}
    \begin{tabular}{l|l|l}
      & Mozilla & Apache \\
      \hline
      \# developers without merging & 9,894 & 40,201 \\
      \# merged developers & 6,810 & 27,627 \\
      \# issue tracker profiles & 271,816 & 139,434 \\
      \% developers in issue tracker & 68 & 44.8 \\
      \% commits from developers in issue tracker & 84 & 83.7 \\
      \% comments from developers & 58.4 & 55.5 \\
    \end{tabular}
  \end{small}
\end{table}

\subsubsection{Handling time zones}

Because Apache and Mozilla relied on SVN and CVS before Git and Mercurial, not all commits have accurate time zones. For some part of the commit history, time zones are simply missing. First, Apache commits imported from SVN could be identified using regular expressions in the commit messages. Second, for each repository, we only consider time zones being accurate starting from the first commits after 2007 that doesn't use UTC as a timezone. This leaves 78.7\% of Mozilla's commit time zones and 59.8\% of Apache's commit time zones usable.

Relying on the identity merging of developer profiles, we inferred time zones for the issue tracker timestamps using the ones used in the commits. For each developer, we listed the time zones they used in commits from all repositories from a given source (Mozilla or Apache). Then, for each timestamp in the issue tracker, a time zone is chosen based on the one used by the developer's previous commit. In total, we could infer time zones for 43.4\% and 55.4\% of all issue and issue comment timestamps for Mozilla, and 57.6\% and 50.7\% for Apache.

\subsubsection{NLP on comments}

Each raw comment was processed with different NLP tools:
\begin{itemize}
    \item Text line classification. We use simple regular expressions to detect lines of text that are automatically generated (only for Bugzilla), quoted or written by the author. In addition, NLoN is used to predict whether each line is natural language or not.
    \item Natural language filtering. All text that wasn't authored by the comment author (generated or quoted text) and all non natural language text (as detected in the previous step) are removed. Lines of text separated by a single line feed are also grouped by paragraphs for Bugzilla. This is done because individual sentences are often split with line feeds.
    \item Emoticon detection. Unicode emojis and emoticons contained in natural language are detected using regular expressions relying on our own R package\footnote{\url{https://github.com/M3SOulu/EmoticonFindeR}}.
    \item Sentence detection. The R package \emph{tokenizers} is used to split each paragraph in sentences.
    \item Sentiment analysis. Both SentiStrength~\cite{sentistrength} and Senti4SD~\cite{Calefato2018} are run on each sentence.
\end{itemize}



\section{Limitations and future work}\label{sec:limitations}

This dataset comes with several limitations that one needs to be aware of. First, it is not exhaustive. This is particularly true for Apache as some projects are still not using Jira and Git. In particular, the famous Apache httpd web server\footnote{\url{https://httpd.apache.org/}} is not included as it still uses Bugzilla and SVN. While it would be possible to extend the dataset to include it in the future, it wouldn't be possible to infer any timezone. Second, timestamp information is incomplete as any source code that was originally versioned with SVN and CVS, and later imported in Git and Mercurial, is missing time zone information. Third, the identity merging performed is quite rudimentary. While the result was initially manually checked to avoid linking distinct developers with more than 100 commits for the previous version of the data (\ie extracted in January 2018 and used in our ICSE paper~\cite{Claes2018ICSE}), it was only re-checked for developers with more than 1000 commits for the updated version due to time constraint.

There are also several limitations regarding the natural language processing of issue comments. First, NLoN's prediction model hasn't been retrained with any Apache data. We are aware of several non natural language Apache comments that are considered as natural language for that reason. The same is true in a lesser extent for Mozilla and this will be improved in the future by adding more manually labelled data to NLoN and re-processing the comments. The description of each issue wasn't included in the NLP pipeline for Apache as it is included as meta-data of the issue, while for Mozilla it is included as the first comment of the issue. In the future, we plan on updating the dataset to include this description field as a comment rather than meta-data.

Finally, as we are still using the dataset, in particular for natural language processing, we will keep on releasing extensions and updates of the dataset in the upcoming years.



\section*{Acknowledgments}

The authors have been supported by Academy of
Finland grants 298020 and 328058.


\bibliographystyle{ACM-Reference-Format.bst}
\bibliography{biblio}

\end{document}